\documentclass[twocolumn,preprintnumbers,prl,floatfix,superscriptaddress,amsmath,amssymb]{revtex4}
\usepackage{epsfig}
\usepackage{subfigure}
\usepackage{amsmath}
\usepackage{amssymb}
\usepackage{setspace}
\usepackage{graphicx}
\usepackage{dcolumn}
\usepackage{bm}
\input epsf

\begin{document}

\author{Juan G. Restrepo}
\email{juanga@neu.edu} \affiliation{Center for Interdisciplinary
Research in Complex Systems, Northeastern University, Boston, MA
02115, USA }

\author{Edward Ott}
\affiliation{ Department of Physics, Department of Electrical and
Computer Engineering, and Institute for Research in Electronics and
Applied Physics, University of Maryland, College Park, Maryland
20742, USA }

\author{Brian R. Hunt}
\affiliation{Department of Mathematics and Institute for Physical
Science and Technology, University of Maryland, College Park,
Maryland 20742, USA}

\date{\today}

\begin{abstract}
We present an analysis of the percolation transition for general
node removal strategies valid for locally tree-like directed
networks. On the basis of heuristic arguments we predict that, if
the probability of removing node $i$ is $p_i$, the network
disintegrates if $p_i$ is such that the largest eigenvalue of the
matrix with entries $A_{ij}(1-p_i)$ is less than $1$, where $A$ is
the adjacency matrix of the network. The knowledge or applicability
of a Markov network model is not required by our theory, thus making
it applicable to situations not covered by previous works. We test
our predicted percolation criterion against numerical results for
different networks and node removal strategies.
\end{abstract}

\title{Weighted percolation on directed networks}


\maketitle

There has been much recent interest in the structure and function of
complex networks \cite{newman1}. One aspect that has received
considerable attention is the resilience of networks to the removal
of some of their nodes
\cite{callaway,cohen,cohen1,cohen2,newman2,moreno,serrano,ser}. This
problem is related, for example, to the robustness of transportation
and information networks to disturbances like random failures or
targeted attacks, or to the resistance of biological networks to the
action of drugs. Another related problem is determining the
threshold for epidemic spreading \cite{grassberger}. An important
objective is to determine what characteristics of the structure of
the network determine the proportion of nodes that can be removed
before the network disintegrates.

A model often considered is one in which nodes are removed from a
network of size $N$ with a uniform probability $p$. In the large $N$
limit, for probabilities less than a critical value $p_c$, there is
a connected component of the network of size of order $N$ (the {\it
giant component}), and for values of $p$ larger than $p_c$, there is
no connected component of size of order $N$
\cite{newman1,callaway,cohen,cohen1,cohen2,newman2,moreno,serrano}.
The critical probability $p_c$ at which this {\it percolation}
transition occurs has been the subject of several theoretical works,
and various approximations have been put forward.

In what follows we define the in and out degrees of a node $i$ by
$d^{out}_i = \sum_{j=1}^N A_{ij}$ and $d^{in}_i = \sum_{j=1}^N
A_{ji}$. Here $A_{ij}$ is the network adjacency matrix; $A_{ij} =1$
if there is a directed link from $i$ to $j$ and $0$ otherwise. If $A
= A^T$ the network is said to be undirected and
$d_{i}^{out}=d_{i}^{in}=d_i$. For undirected, degree uncorrelated
networks (the number of connections per node for neighboring nodes
is not correlated), Cohen et al. have shown \cite{cohen} that the
critical probability is approximately given by
$(1-p_c)\left[(\langle d^2\rangle/\langle d\rangle)-1\right] = 1$,
where $\langle \cdot \rangle$ denotes an average over nodes.
Reference ~\cite{moreno} treats the case of undirected networks with
correlations  for {\it degree Markovian} networks, i.e., networks in
which all nontrivial correlations are captured by the probability
$P( d'| d)$ that a randomly chosen link from a node with degree $d$
is connected to a node with degree $ d'$. Reference~\cite{moreno}
generalizes the result in Ref.~\cite{cohen} and obtains a critical
value of $p_c$ given by
$\addtolength{\belowcaptionskip}{-0.6cm}\addtolength{\abovecaptionskip}{-0.6cm}
(1-p_c)\Lambda = 1,$ where $\Lambda$ is the largest eigenvalue of
the matrix with entries $C_{dd'} = (d'-1) P(d'|d)$. Other works on
undirected networks have extended the Markovian approach to include
the effect of clustering (e.g., often present in social networks);
in particular Ref.~\cite{ser} presents a general analysis for the
case of weak clustering.

Reference~\cite{serrano} studies the percolation transition in
directed degree Markovian networks. The types of components studied
are a {\it strongly connected component} (SCC), defined as a set of
nodes such that every node in the SCC is reachable from any other
node in the SCC by a directed path, its associated {\it
in-component} (IN), defined as the set of nodes from which the SCC
is reachable by a directed path, and {\it out-component} (OUT),
defined as the set of nodes reachable from the SCC by a directed
path. Notice that there might be several such components in a
particular network. Of interest is the largest strongly connected
component which, if its size is of order $N$, is called the {\it
giant strongly connected component}, GSCC. The out and in components
of the GSCC are denoted GOUT and GIN.

It was found in Ref.~\cite{serrano} that as the probability of node
removal $p$ is increased, GSCC, GOUT, and GIN disappear at the same
critical value $p_c$. This value was found to be determined by the
largest eigenvalue of a matrix expressed in terms of $P_o\left({\bf
y}'|{\bf y}\right)$ and $P_b\left({\bf y}'|{\bf y}\right)$ where
${\bf y} = (d_p^{in},d_p^{out},d_p^{bi})$, and $d_p^{in}$,
$d_p^{out}$ and $d_p^{bi}$ are the number of purely incoming, purely
outgoing and bidirectional edges for a given node (i.e., an edge
$A_{ij}=1$ is purely outgoing from node $i$ if $A_{ji}$ = 0).
$P_o\left({\bf y}'|{\bf y}\right)$ and $P_b\left({\bf y}'|{\bf
y}\right)$ are the probabilities of reaching a node of degree ${\bf
y}'$ from a node of degree ${\bf y}$ by following an outgoing and a
bidirectional edge, respectively.

One of our aims in this paper is to remove the need for the
applicability and knowledge of a Markov network model. In order to
do so, in this paper we will focus on a class of directed networks
that are locally tree-like in the sense that they have few short
loops. More precisely, we assume that for each node $i$ and not too
large $L$, the number of {\it different} nodes reachable by paths of
length $L$ or less starting at node $i$ is close to the {\it total}
number of paths of length $L$ or less starting from node $i$. In
particular ($L=2$) we will assume that bidirectional edges are
absent or negligible in number. Under this assumption, ${\bf y} =
(d^{in},d^{out},0)$, and the matrix in Ref.~\cite{serrano} whose
eigenvalue determines the percolation transition reduces to
\begin{equation}\label{chunga2}
\addtolength{\belowcaptionskip}{1cm}\addtolength{\abovecaptionskip}{1cm}
\hat C_{{\bf zz}'} = (d^{out})' P({\bf z}'|{\bf z}),
\end{equation}
where ${\bf z}=(d^{in},d^{out})$. We note that our locally tree-like
condition for directed networks is analogous to assuming negligible
clustering.

In many situations the node removal probability is not a constant.
For example, airports might have different security measures, or
differ in their vulnerability to an attack or weather related
shutdown due to their geographical location. Also, we have noted
recently \cite{imp} that a measure of the dynamical importance of
nodes is proportional to $v_i u_i$, where $u$ and $v$ are the right
and left eigenvectors corresponding to the largest eigenvalue
$\lambda$ of the adjacency matrix of the network, $A$: $Au = \lambda
u$, $v^T A = \lambda v^T$. Thus, a potential removal strategy (to be
used in an example later in this paper) is that in which nodes are
removed based on the value of $v_i u_i$. More generally, we would
like to study the effect of node removal strategies that assign a
probability $p_i$ to the removal of node $i$, and we refer to this
problem as {\it weighted percolation}. (Other previously considered
possibilities are that highly connected nodes are preferentially
removed from the network, e.g., $p_i$ is proportional to the degree
of node $i$~\cite{cohen1}, or to a power of the degree of node $i$
\cite{cohen2}.)

Our objective is to present a simple heuristic method for treating
{\it general} weighted percolation removal strategies (i.e., general
$p_i$) on directed networks without the need for a Markovian network
model. While we do not use the Markovian assumption nor a specific
node removal probability, we assume a locally tree-like network
structure (we will discuss the validity of this assumption below),
and we also require knowledge of the network adjacency matrix $A$.
We find that the network disintegrates, as defined by the
disappearance of the giant connected components, when the node
removal strategy is such that the largest eigenvalue $\hat{\lambda}$
of the matrix $\hat A$ with entries $\hat A_{ij} = A_{ij}(1 - p_i)$
is less than $1$.

To obtain the above result we adapt the mean field arguments given
for example in \cite{newman2} and \cite{moreno} to our case.
Consider first the disappearance of the giant in-component GIN. Let
$\eta_i$ be the probability that node $i$ is not in the giant
in-component GIN. Node $i$ is not in GIN either if it has been
removed (with probability $p_i$), or if it has not been removed and
none of its out-links point to nodes in GIN. Consider two such nodes
$j_1$ and $j_2$ that $i$ points to. We argue that it is reasonable
to make the approximation that whether $j_1$ belongs to GIN is
independent of whether $j_2$ belongs to GIN. Whether $j_1$ is in GIN
depends on whether the nodes it points to are in GIN, which depends
on the nodes they point to, and so on. Our locally tree-like
assumption implies that the nodes that can be reached from $j_1$ by
a short path are essentially independent of the nodes that can be
reached from $j_2$ by a short path. Based on this independence
assumption, $\eta_i$ is given by (recall that $A_{ij} = 0$ or $1$) $
\eta_i = p_i +(1-p_i)\prod_{j=1}^N (\eta_j)^{A_{ij}} $. This
equation always has the trivial solution $\eta_i = 1$. The presence
of a giant in-component requires a solution for which the expected
size $s = N - \sum_{j=1}^N \eta_i$ is positive. Setting $\eta_i =
e^{-z_i}$ and assuming $0 \leq z_i$ and $\sum_{j=1}^N A_{ij} z_j \ll
1$, we obtain the approximation $ z_i = \sum_{j=1}^N A_{ij}(1-p_i)
z_j$. When $p_i = 1$ (i.e., all nodes are removed), the only
solution is the trivial solution $z_i = 0$.  As we decrease the
$p_i$'s, a nontrivial solution (corresponding to a giant
in-component) first appears when the largest eigenvalue $\hat
\lambda$ of the matrix $\hat A$ with entries $\hat A_{ij} =
A_{ij}(1-p_i)$ is $1$. Note that, as required, we can satisfy
$\eta_i\leq 1$ since the components $z_i$ of the eigenvector
corresponding to $\hat \lambda$ are, by the Frobenius theorem
\cite{siamlambda}, nonnegative. Applying the same reasoning to the
out-component GOUT we find that it appears when the largest
eigenvalue of the matrix with entries $B_{ij}=A_{ji}(1-p_j)=(\hat
A^T)_{ij} $ is $1$. Since the transpose of $\hat A$ and $\hat A$
have the same spectrum, the giant in and out-components appear
simultaneously.

The above can also be understood by the following heuristic
argument. Our previous discussion applies not only to GOUT (GIN),
but more generally to sets generated by repeatedly following
outgoing (incoming) links starting from a given node. Therefore, one
can estimate the size of such sets and locate the transition as the
point at which one of them has macroscopic size. In doing so, it is
essential not to overcount the number of nodes, and it is here where
our assumption of locally tree-like network structure allows us to
simplify the problem. The number of directed paths of length $m$
starting from node $i$ can be estimated using this assumption, for
not too large $m$, as the sum of the components of the vector ${\hat
A}^m e^i$, where $e^i$ is the unit vector for coordinate $i$. If the
largest eigenvalue of $\hat A$ is larger than $1$, the number of
paths of length $m$ grows exponentially with $m$ for some starting
node $i$. Under our assumptions, these paths traverse different
nodes, and thus the out-component of $i$ has large size, in
agreement with our previous result.

In order to motivate the locally tree-like assumption, we consider
the illustrative case of uncorrelated networks. (While our concern
in this paper is not with uncorrelated networks, they, nevertheless,
provide a useful example of why the locally tree-like assumption
might be valid in some cases.) In particular, we estimate the
fraction of bidirectional edges. The probability $p_{ij}$ that nodes
$i$ and $j$ share a bidirectional edge is given by
$p_{ij}=d_i^{out}d_i^{in}d_j^{in}d_j^{out}/(N^2\langle d
\rangle^2)$, where, since $\langle d^{in}\rangle = \langle
d^{out}\rangle$, we use $\langle d\rangle$ to denote either of these
averages.
 In the case that the degrees at nodes $i$ and $j$ are uncorrelated,
 on average, $\langle p_{ij}\rangle =
\left(\langle d_i^{out}d_i^{in}\rangle/(N\langle d \rangle)\right)^2
\approx \left(\lambda /N\right)^2,$ where we have used the mean
field approximation for the maximum eigenvalue $\lambda$ of $A$ (see
below). Not too far above the percolation transition $\lambda$ is of
order $1$, and we thus expect the locally tree-like assumption to be
valid.

We next discuss how our results compare to those of
Ref.~\cite{serrano} in the case of negligibly few bidirectional
edges (i.e., Eq.(\ref{chunga2})). If $p_i = p$, our result for the
critical probability $p_c$ reduces to $(1-p_c)\lambda=1$, where
$\lambda$ is the largest eigenvalue of the adjacency matrix $A$. If
the network is degree Markovian, and we let $\psi_{\bf z}^{(m)}$ be
the average number of directed paths of length $m$ starting from
nodes of degree ${\bf z}$, we have $\psi_{{\bf z}}^{(m+1)} =
d^{out}\sum_{{\bf z}'}P({\bf z}'|{\bf z})
 \psi_{{\bf z}'}^{(m)}$. Since, for large $m$, the number of paths of length $m$ grows like
$\lambda^m$, we associate to the previous equation the eigenvalue
problem $\lambda_M\psi_{{\bf z}} = d^{out}\sum_{{\bf z}'}P({\bf
z}'|{\bf z})\psi_{{\bf z}'}$, where $\lambda_M$ is the {\it
Markovian approximation} to $\lambda$. The previous result agrees
with Eq.~(\ref{chunga2}) [the matrices $d^{out} P( {\bf z}'|{\bf
z})$ and $(d^{out})' P({\bf z}'|{\bf z})$ have the same spectrum].
We note that, in the absence of degree-degree correlations, we have
$P({\bf z}'|{\bf z}) = d^{in} P({\bf z}')/\langle d\rangle$, which
when inserted in the eigenvalue equation yields the {\it mean field
approximation} for the eigenvalue, $\lambda_{mf} = \langle
d^{in}d^{out}\rangle/\langle d\rangle$.

We now illustrate our theory with two numerical examples. The first
example ({\it Example 1}) illustrates the flexibility of our
approach to address various weighted percolation node removal
strategies, while the second example ({\it Example 2}) illustrates
the point that our approach does not require the knowledge or
applicability of a Markov network model.
\begin{figure}[h]
\centering \epsfig{file = 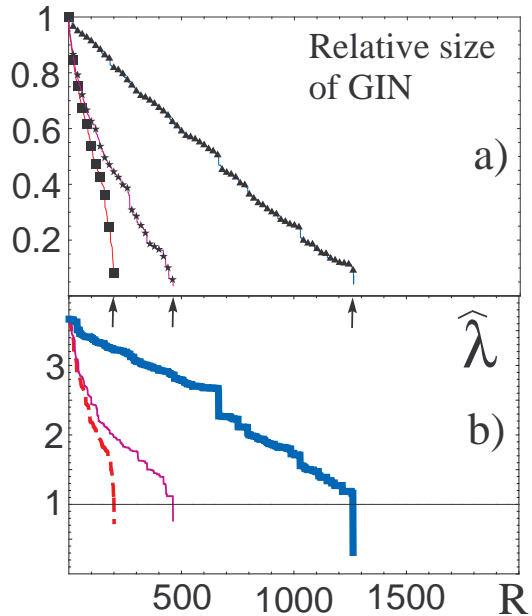, clip =  ,width=0.9\linewidth }
\addtolength{\belowcaptionskip}{-0.4cm}\addtolength{\abovecaptionskip}{-0.5cm}
\caption{(a) Ratio of the largest connected component to the number
of remaining nodes, and (b) largest eigenvalue of the adjacency
matrix of the network after node removal as a function of the number
of removed nodes $R$, when nodes are removed randomly (triangles and
thick solid line in a) and b) respectively), in order of decreasing
$d_i^{out}d_i^{in}$ (stars and thin solid line), and in order of
decreasing dynamical importance (boxes and dashed line)(see text).
The percolation transition occurs when $\hat \lambda = 1$ [indicated
by the vertical arrows in (a)].} \label{fi0}
\end{figure}

{\it Example 1.} For simplicity, we consider uncorrelated random
networks with degree distributions $P(d^{in},d^{out})$ in which
$d^{in}$ and $d^{out}$ are independent and have the same
distribution $\tilde P(d)$, that is, $P(d^{in},d^{out})=\tilde
P(d^{in})\tilde P(d^{out})$. We use a generalization of the method
in Ref.~\cite{chung} in order to generate networks with a power law
degree distribution, $\tilde P(d)\propto d^{-\gamma}$. We choose the
sequence of expected degrees ${\tilde d^{in}_i} =
c(i+i_0-1)^{-1/(\gamma-1)}$ for the in-degrees, and a random
permutation of this sequence for the out-degrees, where
$i=1,\dots,N$, and $c$ and $i_0$ are chosen to obtain a desired
maximum and average degree. Then, the adjacency matrix is
constructed by setting $A_{ij} = 1$ for $i\neq j$ with probability
$\tilde{d}_i^{out}\tilde{d}_j^{in}/(N\langle d \rangle)$ and zero
otherwise ($A_{ii} = 0$). The ensemble expected value of the
resulting network degree distribution is given by
$P(d^{in},d^{out})$. (Note that we assume
$\tilde{d}_i^{out}\tilde{d}_j^{in}< N\langle d \rangle$.) In
Fig.~\ref{fi0}(a) we show, for a $N = 2000$ scale free network with
exponent $\gamma = 2.5$ and $\langle d\rangle =3$, the size of GIN
as a function of the number of removed nodes $R$, when nodes are
removed in order of decreasing $d_i^{in} d_i^{out}$ (thick solid
line), decreasing dynamical importance \cite{imp} ($v_iu_i/v^Tu$)
(thin solid line), and randomly (dashed line) \cite{foot}. The
removal probabilities in the first two cases are given by $p_i = 1$
if $i\in S$ and $0$ otherwise for a subset $S$ of nodes. (For
example, in the first case $S=\{i: d_i^{in} d_i^{out} > d_*^2 \}$,
and $\hat A$ reduces to the matrix obtained by removal of all nodes
in $A$ for which $d_i^{in} d_i^{out}> d_*^2$.) We also show in
Fig.~\ref{fi0}(b) the largest eigenvalue $\hat \lambda$ of $\hat A$
which in this case is equivalent to the adjacency matrix of the
network resulting from the removal of the nodes. We observe for all
three cases that the network disintegrates, as predicted, when $\hat
\lambda = 1$ (indicated with the vertical arrows in
Fig.~\ref{fi0}(b)). The number of removed nodes necessary to
disintegrate the network is less when removing nodes by dynamical
importance. Removal of nodes by $d_i^{in} d_i^{out}$ requires
somewhat more nodes to disintegrate the network, while random
removal requires removal of the most.
\begin{figure}[h]
\centering \epsfig{file = 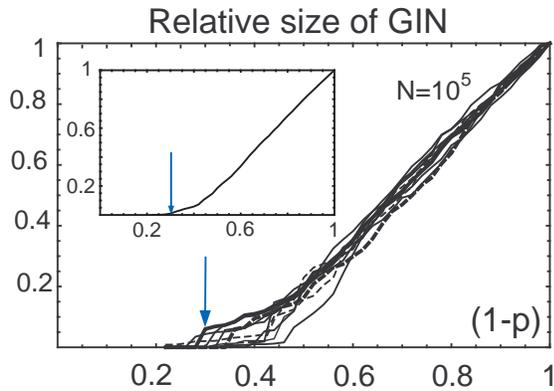, clip = ,width=0.90\linewidth }
\addtolength{\belowcaptionskip}{-0.5cm}\addtolength{\abovecaptionskip}{-0.4cm}
\caption{ Normalized size of the GIN as a function of the number of
remaining nodes, $1-p$, for {\it Example 2}. The plot shows ten
different realizations of the node removal process, and the inset
shows their mean } \label{remo}
\end{figure}

The simplest Markovian network assumption is that correlations
between connected nodes depend solely on the degree of these nodes
and not on any other property the nodes might have. While this is a
useful analytical framework, probably applicable to many cases, it
is also likely to fail in other cases (e.g., \cite{kim}). In our
next example we will consider a network in which the node statistics
depends on variables additional to the degrees and for which,
therefore, a degree based Markovian approximation is not a priori
expected to apply.

{\it Example 2.} We start with a network generated as in {\it
Example 1} with $N=10^5$, $\gamma = 2.5$, and $\langle d \rangle =
3$. Then, we first specify a division of the nodes in the network
into two groups of the same size, $A$ and $B$, ($A\bigcup B =
\{1,2,\dots,N\}$, $A\bigcap B = \emptyset$). We define a measure of
the degree-degree correlations $\rho = \langle
d_i^{in}d_j^{out}\rangle_e/\langle d_i\rangle_e^2$, where
$\langle\dots\rangle_e$ denotes an average over edges, $\langle
Q_{ij}\rangle_e \equiv \sum_{i,j}A_{ij}Q_{ij}/\sum_{i,j}A_{ij}$. The
following (an adaptation of the method in Ref.~\cite{mixing}) is
repeated until the network has the desired amount of degree-degree
correlations as evidenced in the value of $\rho$: Two edges are
chosen at random, say connecting node $i$ to node $j$ and node $n$
to node $m$. If $i,j,n,m$ are all in $A$, the edges are replaced
with two edges connecting node $i$ to node $m$ and node $n$ to node
$j$ if $(d_n^{in}d^{out}_m +
d_i^{in}d^{out}_j-d_n^{in}d^{out}_j-d_i^{in}d^{out}_m) < 0$. If
$i,j,n,$ and $m$ are all in $B$, the edges are replaced with two
edges connecting node $i$ to node $m$ and node $n$ to node $j$ if
$(d_n^{in}d^{out}_m +
d_i^{in}d^{out}_j-d_n^{in}d^{out}_j-d_i^{in}d^{out}_m) > 0$.
Otherwise the edges are unchanged. The effect of this division is to
create two subnetworks, $A$ and $B$, with positive and negative
degree-degree correlations. Starting from such a network, we
successively remove a randomly chosen node and compute the size of
the GIN relative to its initial size. In Fig.~\ref{remo} we plot
this normalized size of the GIN as a function of the fraction of
remaining nodes for ten realizations of the node removal sequence.
The vertical arrow represents the prediction from the eigenvalue.
Although the transition points of individual realizations have some
spread, the arrow gives a good approximation of their mean (see
inset).

We now discuss the advantages and disadvantages of the eigenvalue
approach when compared to the Markov approximation. As opposed to
the Markov approximation, the eigenvalue approximation allows the
easy treatment of general node removal strategies (`weighted
percolation'). Furthermore, it does not require the assumption that
the node correlations depend only on their degree and are only to
nearest neighbors. In addition, the construction of the matrix
$d^{out} P( {\bf z}'|{\bf z})$ and the determination of its largest
eigenvalue is in some cases harder than the direct determination of
the largest eigenvalue of the adjacency matrix $A$. On the other
hand, in many cases the adjacency matrix of the network is not
known, and local sampling methods from which an approximation to the
matrix $d^{out} P( {\bf z}'|{\bf z})$ can be constructed must be
used. Additionally, the eigenvalue approach is valid only when the
network has locally tree-like structure.

In conclusion, we have presented a simple eigenvalue-based criterion
for percolation on directed networks. Our method should be viewed as
complementary to previous studies in that it does not require
knowledge or applicability of a Markov network model and can treat
general node removal strategies, but requires knowledge of the
network adjacency matrix $A$ and only applies when the network has
locally tree-like structure.

This work was supported by ONR (Physics), by the NSF (PHY 0456240
and DMS 0104-087), and by AFOSR FA95500410319.
\addtolength{\parskip}{-0.6cm}

\begin{spacing}{1.0}
\bibliographystyle{plain}

\begin{thebibliography}{99}

\bibitem{newman1} M.E.J. Newman, SIAM Review {\bf 45}, 167 (2003);
A.-L. Barab\'{a}si, and R. Albert, Rev. Mod. Phys. {\bf 74}, 47
(2002); S. Boccaletti {\it et al.}, Phys. Rep. {\bf 424}, 175
(2006).



\bibitem{callaway} D. S. Callaway {\it et al.}, Phys. Rev. Lett. {\bf 25}, 5468 (2000).

\bibitem{cohen} R. Cohen {\it et al.}, Phys. Rev. Lett. {\bf
85}, 4626 (2000).

\bibitem{cohen1} R. Cohen, S. Havlin, and D. ben-Avraham, Phys. Rev. Lett. {\bf
91}, 247901 (2003).

\bibitem{cohen2} L. K. Gallos {\it et al.}, Phys. Rev. Lett. {\bf
94}, 188701 (2005).

\bibitem{newman2} M. E. J. Newman, Phys. Rev. Lett. {\bf 89} 208701
(2002).

\bibitem{moreno} A. V\'azquez and Y. Moreno, Phys. Rev. E {\bf 67}, 015101(R)
(2003).

\bibitem{serrano} M. Bogu\~n\'a and M. A. Serrano, Phys. Rev. E {\bf 72},
016106 (2003).


\bibitem{ser} M. A. Serrano and M. Bogu\~n\'a, Phys. Rev. E {\bf 74}, 056115 (2006).


\bibitem{grassberger} P. Grassberger, Math. Biosci. {\bf 63} 157
(1983). This paper gives a mapping of the epidemic problem onto the
percolation problem.


\bibitem{imp} J. G. Restrepo, E. Ott, and
B. R. Hunt, Phys. Rev. Lett. {\bf 97}, 094102 (2006).

\bibitem{siamlambda} C. R. MacCluer, SIAM Review {\bf 42}, 487 (2000).

\bibitem{chung} F. Chung, L. Lu and V. Vu, Proc. Natl. Acad. Sci., {\bf 100}, 6313  (2003).


\bibitem{foot} Note that the first two removal criteria are related in that
$d_i^{out} d_i^{in}/ \sum (d_i^{out} d_i^{in})$  can be viewed as a
mean field prediction of the dynamical importance.


\bibitem{kim} J. W. Kim, B. R. Hunt and E. Ott, Phys. Rev. E {\bf 66}, 046115 (2002).

\bibitem{mixing} M. E. J. Newman, Phys. Rev. E {\bf 67}, 026126
(2003).

\end{thebibliography}

\end{spacing}

\end{document}